\documentclass[12pt]{article}
\usepackage{epsf,amssymb,amsmath,latexsym}
\title{Lyapunov exponents at anomalies of SL$(2,\RR)$-actions}

\author{Hermann Schulz-Baldes
\\
\\
{\small Mathematisches Institut, Universit\"at
Erlangen-N\"urnberg, Germany}
}

\date{ }

\newtheorem{defini}{Definition}
\newtheorem{proposi}{Proposition}
\newtheorem{lemma}{Lemma}

\newcommand{\NN}{{\mathbb N}}
\newcommand{\RR}{{\mathbb R}}

\newcommand{\ZZ}{{\mathbb Z}}

\newcommand{\hth}{{\hat{\theta}}}

\newcommand{\pp}{{\bf p}}
\newcommand{\EE}{{\bf E}}

\newcommand{\Ss}{{\cal S}}
\newcommand{\Oo}{{\cal O}}
\newcommand{\Tr}{\mbox{\rm Tr}}

\newcommand{\Ll}{{\cal L}}

\begin{document}

\maketitle

%%%%%%%%%%%%%%%%%%%%%%%%%%%%%%%%%%%%%%%%%%%%%%%%%%%%
\begin{abstract}
Anomalies are known to appear in the perturbation theory for the 
one-dimensional Anderson model. A systematic approach to anomalies 
at critical points of products of random matrices is developed,
classifying and analysing their possible types. The associated
invariant measure is calculated formally. For an anomaly of so-called 
second degree, it is given by the groundstate of a certain
Fokker-Planck equation on the unit circle. 
The Lyapunov exponent is calculated
to lowest order in perturbation theory with rigorous control of the
error terms. 
\end{abstract}
%%%%%%%%%%%%%%%%%%

\vspace{.5cm}

%%%%%%%%%%%%%%%%%%%%%%%%%%%%%%%%%%%%%%%%%%
\section{Introduction}

Anomalies in the perturbative calculation of the Lyapunov exponent
and the density of states were first found and analysed by Kappus and Wegner
\cite{KW} when they
studied the center of the band in a one-dimensional Anderson
model. Further anomalies, albeit in higher order perturbation theory,
were then treated by Derrida and Gardner \cite{DG} as well as Bovier
and Klein \cite{BK}. More recently,
anomalies also appeared in the study of random polymer models
\cite{JSS}. Quite some effort has been made to understand anomalies in
the particular case of the Anderson model also from a more mathmatical
point of view \cite{CK,SVW}. However, Campanino
and Klein \cite{CK} need to suppose decay estimates on the
charcacteristic function of the random potential, and Shubin, 
Vakilian and Wolff \cite{SVW} appeal to rather complicated techniques
from harmonic analysis (allowing only to give the correct scaling of
the Lyapunov exponent, but not a precise perturbative formula for it).

\vspace{.2cm}

It is the purpose of this work to present a more conceptual approach
to anomalies of products of random matrices. 
In fact, various types may appear and only those of
second degree (in the sense of the definition below) seem to have 
been studied previously. Indeed, this is the most difficult and interesting
case to analyse, and the main insight of the present work is to
exhibit an associated Fokker-Planck operator, the spectral gap of
which is ultimately responsible for the positivity of the Lyapunov
exponent. In the special case of the Anderson model, a related operator 
already appeared in \cite{BK}. Here it
is, however, possible to circumvent the spectral analysis of the
Fokker-Planck
operator and prove the asymptotics of the Lyapunov exponent more
directly ({\it cf.} Section~\ref{sec-second}). The other cases of
various first degree anomalies are more elementary to
analyse. Examples for different types of anomalies are given in
Section~\ref{sec-examples}. 

\vspace{.2cm}

\noindent {\bf Acknowledgement:} This is a contribution to the conference
procedeings of OTAMP2004 held in Bedlewo in June 2004.
First of all, I would like to thank the
organizers for the invitation to the wonderful and inspiring
conference in the convivial atmosphere of Bedlewo, allowing for
quite unexpected encounters; second of all, many thanks to my
new collegues Andreas Knauf and Hajo Leschke for a few 
patient discussions on the matters of the present work.

%%%%%%%%%%%%%%%%%%%%%%%%%%%%%%%%%%%%%%%%%%
\section{Definition of anomalies}

Let us consider families
$(T_{\lambda,\sigma})_{\lambda\in\RR,\sigma\in\Sigma}$ of
matrices in SL$(2,\RR)
=\{T\in \mbox{Mat}_{2\times 2}(\RR)\,|\,\det(T)=1\}$ depending on 
a random variable $\sigma$ in some probability space
$(\Sigma,{\bf p})$ as well as a real coupling parameter
$\lambda$. In order to avoid technicalities, we suppose that ${\bf p}$
has compact support. The dependence on $\lambda$ is supposed to be
smooth. The expectation value w.r.t. ${\bf p}$ will be denoted by
$\EE$. 

%%%%%%%%%%%%%%%%%%%%%%%%%%%%%%%%%%%%%%%%%%
\begin{defini}
\label{def-critical}
The value $\lambda=0$ is anomaly of first order of the family 
$(T_{\lambda,\sigma})_{\lambda\in\RR,\sigma\in\Sigma}$ if for all
$\sigma\in\Sigma$:

\begin{equation}
\label{eq-sign}
T_{0,\sigma}\;=\;\pm\,{\bf 1}
\mbox{ , }
\end{equation}

\noindent with a sign that may depend on $\sigma\in\Sigma$. In order to
further classify the anomalies and for later
use, let us introduce $P_\sigma,Q_\sigma\in\mbox{\rm sl}(2,\RR)$
by

\begin{equation}
\label{eq-expan}
MT_{\lambda,\sigma}M^{-1}
\;=\;
\pm\,\exp\left(\lambda\,P_\sigma\;+\;\lambda^2\,Q_\sigma
\;+\;\Oo(\lambda^3)
\right)
\;,
\end{equation}

\noindent where $M\in \mbox{\rm SL}(2,\RR)$ is a $\lambda$- and
$\sigma$-independent
basis change to be chosen later. An
anomaly is said to be of first degree if $\EE(P_\sigma)$ is
non-vanishing, and then it is called elliptic if $\det(\EE(P_\sigma))>0$,
hyperbolic if
$\det(\EE(P_\sigma))<0$ and parabolic if $\det(\EE(P_\sigma))=0$. 
Note that all these notions are
independent of the choice of $M$.

\vspace{.2cm}

If
$\EE(P_\sigma)=0$, but
the variance of $P_\sigma$ is non-vanishing, then an
anomaly is said to be of second degree. 

\vspace{.2cm}

Furthermore, for $k\in\NN$, set
$\hat{\sigma}=(\sigma(k),\ldots,\sigma(1))\in\hat{\Sigma}=\Sigma^{\times
k}$, as well as $\hat{\pp}=\pp^{\times k}$ 
and $T_{\lambda,\hat{\sigma}}=T_{\lambda,\sigma(k)}\cdots
T_{\lambda,\sigma(1)}$. Then $\lambda=0$ is anomaly of $k$th
order of the family 
$(T_{\lambda,\sigma})_{\lambda\in\RR,\sigma\in\Sigma}$ if the family
$(T_{\lambda,\hat{\sigma}})_{\lambda\in\RR,\hat{\sigma}\in\hat{\Sigma}}$
has an anomaly of first order at $\lambda=0$ in the above sense. The
definitions of degree and nature transpose to $k$th order anomalies.
\end{defini}
%%%%%%%%%%%%%%%%%%%%%%%%%%%%%%%%%%%%%%%%%%

As is suggested in the definition and will be further explained below, we
may (and will) restrict ourselves to the analysis of anomalies of first
order. In the examples, however, anomalies of higher order do appear
and can then be studied by the present techniques
({\it cf.} Section \ref{sec-examples}). Furthermore,
by a change of variables in $\lambda$, anomalies of
degree higher than 2 can be analysed like an anomaly of second degree.

\vspace{.2cm}

Anomalies are particular cases of so-called critical points studied in
\cite{JSS,SSS}, namely $\lambda=0$ is by definition a critical point of the
family 
$(T_{\lambda,\sigma})_{\lambda\in\RR,\sigma\in\Sigma}$ if for all
$\sigma,\sigma'\in\Sigma$:

\begin{equation}
\label{eq-critical}
[\,T_{0,\sigma},T_{0,\sigma'}]\;=\;0
\mbox{ , }
\qquad
\mbox{and}
\qquad
|\Tr(T_{0,\sigma})|\;<\;2
\;\mbox{ or }\;T_{0,\sigma}\;=\;\pm\,{\bf 1}
\mbox{ . }
\end{equation}

\noindent Critical points appear in many applications like the Anderson
model and the random polymer model. 
In these situations anomalies appear for
special values of the parameters, such as the energy or the coupling
constant, {\it cf.} Section~\ref{sec-examples}. 

\vspace{.2cm}

%%%%%%%%%%%%%%%%%%%%%%%%%%%%%%%%%%%%%%%%%%
\section{Phase shift dynamics}

The bijective action $\Ss_T$ of a matrix $T\in\mbox{SL}(2,\RR)$ on 
$S^1= [0,2\pi)$ is given by

\begin{equation}
\label{eq-action}
e_{\Ss_T(\theta)}
\;=\;
\frac{Te_{\theta}}{
\|Te_{\theta}\|}
\;,
\qquad
\;\;\;\;\;
e_\theta
\;=\;
\left(
\begin{array}{cc} \cos(\theta) \\ \sin(\theta)
\end{array}
\right)
\mbox{ , }
\;\;
\theta\in [0,2\pi)
\mbox{ . }
\end{equation}

\noindent  This defines a group action, namely
$\Ss_{TT'}=\Ss_{T}\Ss_{T'}$. In particular the map $\Ss_T$ is invertible
and $\Ss_T^{-1}=\Ss_{T^{-1}}$. Note that this is actually an action on $\RR
$P$(1)$, and  $S^1$ appears as a double cover here.
In order to shorten notations, we write 

$$
\Ss_{\lambda,\sigma}
\;=\;
\Ss_{MT_{\lambda,\sigma}M^{-1}}
\;.
$$

\noindent Next we need to iterate this dynamics.
Associated to a given semi-infinite code
$\omega=(\sigma_n)_{n\geq 1}$ with $\sigma_n\in\Sigma$ is
a sequence of matrices $(T_{\lambda,\sigma_n})_{n\geq 1}$.  Codes are
random and chosen independently according to the product law ${\bf
p}^{\otimes \NN}$. Averaging w.r.t. ${\bf p}^{\otimes \NN}$ is also 
denoted by $\EE$. Then one defines iteratively for $N\in\NN$

\begin{equation}
\label{eq-randomdyn}
\Ss^N_{\lambda,\omega}(\theta)
\;=\;
\Ss_{{\lambda,\sigma_N}}
\bigl(\Ss^{N-1}_{\lambda,\omega}(\theta)\bigr)
\;,
\qquad
\Ss^0_{\lambda,\omega}(\theta)\;=\;\theta
\;.
\end{equation}

\noindent  This is a discrete time random dynamical
system on $S^1$.
Let us note that at an anomaly of first order, one has
$\Ss_{\lambda,\sigma}(\theta)=\theta+\Oo(\lambda)$ or
$\Ss_{\lambda,\sigma}(\theta)=\theta+\pi+\Oo(\lambda)$ depending on the
sign in (\ref{eq-sign}). As all the functions appearing below will be
$\pi$-periodic we can neglect the summand $\pi$, meaning that we may
suppose that there is a sign + in (\ref{eq-sign}) for all $\sigma$ (this
reflects that the action is actually on projective space). 

\vspace{.2cm}

In order to do perturbation theory in $\lambda$, we need some notations.
Introducing the unit vector
$v=\frac{1}{\sqrt{2}}\left(\begin{array}{c}
1\\-\imath\end{array}\right)$, we define the first order polynomials in
$e^{2\imath \theta}$

$$
p_\sigma(\theta)
\;=\;
\Im m \left(\frac{\langle v|P_\sigma|e_\theta\rangle}{\langle
v|e_\theta\rangle}
\right)
\;,
\qquad
q_\sigma(\theta)
\;=\;
\Im m \left(\frac{\langle v|Q_\sigma|e_\theta\rangle}{\langle
v|e_\theta\rangle}
\right)
\;,
$$

\noindent as well as

$$
\alpha_\sigma
\;=\;
\langle v|P_\sigma|v\rangle\;,
\qquad
\beta_\sigma
\;=\;
\langle \overline{v}|P_\sigma|v\rangle\;.
%\qquad
%\alpha'_\sigma
%\;=\;
%\langle v|Q_\sigma|v\rangle\;,
%\qquad
%\beta'_\sigma
%\;=\;
%\langle v|Q_\sigma|\overline{v}\rangle
%\;,
$$

\noindent Hence
$p_\sigma(\theta)=\Im m(
\alpha_\sigma-\beta_\sigma\,e^{2\imath\theta})$.
Now starting from the identity 

$$
e^{2\imath \Ss_{\lambda,\sigma}(\theta)}
\;=\;
\frac{\langle v|MT_\sigma M^{-1}|e_\theta\rangle}{\langle 
\overline{v}|MT_\sigma
M^{-1}|e_\theta\rangle}
\;,
$$

\noindent the definition (\ref{eq-expan}) and the identity
$\langle v|e_\theta\rangle=\frac{1}{\sqrt{2}}\;e^{\imath\theta}$, one can
verify that 

$$
\Ss_{\lambda,\sigma}(\theta)
\;=\;
\theta\,+\,
\Im m 
\left(
\lambda\,\frac{\langle v|P_\sigma|e_\theta\rangle}{\langle
v|e_\theta\rangle}
\,+\,
\frac{\lambda^2}{2}\,\frac{\langle v|2\,Q_\sigma+P^2_\sigma|e_\theta\rangle}{\langle
v|e_\theta\rangle}
\,-\,
\frac{\lambda^2}{2}\,\frac{\langle v|P_\sigma|e_\theta\rangle^2}{\langle
v|e_\theta\rangle^2}
\right)
\,+\,\Oo(\lambda^3)
\;.
$$

\noindent As  one readily verifies that 

$$
P^2_\sigma
\;=\;
-\det(P_\sigma)\; {\bf 1}\;,
\qquad
\Im m 
\left(
\frac{\langle v|P_\sigma|e_\theta\rangle^2}{\langle
v|e_\theta\rangle^2}
\right)
\;=\;
-\,p_\sigma(\theta)\,
\partial_\theta\, p_\sigma(\theta)
\;,
$$

\noindent it follows that

\begin{equation}
\label{eq-polydef}
\Ss_{\lambda,\sigma}(\theta)
\;=\;
\theta\,+\,\lambda\,p_\sigma(\theta)\,+\,\lambda^2\,q_\sigma(\theta)
\,+\,\frac{1}{2}\,\lambda^2\, p_\sigma(\theta)\,
\partial_\theta \,p_\sigma(\theta)
\,+\,\Oo(\lambda^3)
\;.
\end{equation}

\noindent Finally let us note that

\begin{equation}
\label{eq-polyinv}
\Ss_{\lambda,\sigma}^{-1}(\theta)
\;=\;
\theta\,-\,\lambda\,p_\sigma(\theta)\,-\,\lambda^2\,q_\sigma(\theta)
\,+\,\frac{1}{2}\,\lambda^2\, p_\sigma(\theta)\,\partial_\theta
p_\sigma(\theta)
\,+\,\Oo(\lambda^3)
\;,
\end{equation}

\noindent as one verifies immediately because 
$\Ss_{\lambda,\sigma}(\Ss_{\lambda,\sigma}^{-1}(\theta))
=\theta+\Oo(\lambda^3)$, or can deduce directly just as above from
the identity
$\exp\left(\lambda\,P_\sigma+\lambda^2\,Q_\sigma+\Oo(\lambda^3)
\right)^{-1}=
\exp\left(-\lambda\,P_\sigma-\lambda^2\,Q_\sigma+\Oo(\lambda^3)
\right)$.

%%%%%%%%%%%%%%%%%%%%%%%%%%%%%%%%%%%%%%%%%%
\section{Formal perturbative formula for the invariant measure}
\label{sec-invariant}

For each $\lambda$, the family
$(MT_{\lambda,\sigma}M^{-1})_{\sigma\in\Sigma}$
and the probability ${\bf p}$ define an
invariant probability measure $\nu_\lambda$ on $S^1$
by the equation

\begin{equation}
\label{eq-invariant}
\int d\nu_\lambda(\theta)\,f(\theta)
\;=\;
\EE\,
\int d\nu_\lambda(\theta)\,f(\Ss_{\lambda,\sigma}(\theta))
\mbox{ , }
\qquad
f\in C(S^1)
\mbox{ . }
\end{equation}

\noindent Furstenberg proved that this invariant measure is unique whenever
the Lyapunov exponent of the associated 
product of random matrices (discussed below) 
is positive ({\it e.g.} \cite{BL}) and in
this situation $\nu_\lambda$ is also known to be H\"older continuous, so,
in particular, it does not contain a point component.
For the study of the invariant measure at an anomaly of order $k$, it is
convenient to iterate (\ref{eq-invariant}):

$$
\int d\nu_\lambda(\theta)\,f(\theta)
\;=\;
\EE\,
\int d\nu_\lambda(\theta)\,f(\Ss^k_{\lambda,\omega}(\theta))
\mbox{ , }
\qquad
f\in C(S^1)
\mbox{ . }
$$

\noindent Replacing $(\Sigma,\pp)$ by $(\hat{\Sigma},\hat{\pp})$
therefore shows that the families 
$(T_{\lambda,\sigma})_{\lambda\in\RR,\sigma\in\Sigma}$ and
$(T_{\lambda,\hat{\sigma}})_{\lambda\in\RR,\hat{\sigma}\in\hat{\Sigma}}$
have the same invariant measure. Hence it is sufficient to study 
anomalies of first order.

\vspace{.2cm}

The aim of this section is to present
a formal perturbative expansion of the invariant measure 
under the hypothesis that it is absolutely
continuous, that is
$d\nu_\lambda(\theta)=\rho_\lambda(\theta)
\frac{d\theta}{2\pi} $ with $\rho_\lambda=\rho_0+\lambda\rho_1
+\Oo(\lambda^2)$. Then (\ref{eq-invariant})
leads to

\begin{equation}
\label{eq-invariant2}
\EE
\left(
\partial_\theta \Ss_{\lambda,\omega}^{-1}(\theta)
\rho_\lambda(\Ss^{-1}_{\lambda,\omega}(\theta))
\right)
\;=\;
\rho_\lambda(\theta)
\;,
\end{equation}

\noindent which with equation (\ref{eq-polyinv}) gives

$$
\rho_\lambda
\,-\,\lambda\,\partial_\theta
\Bigl(\EE(p_\sigma)\rho_\lambda\Bigr)
\,+\,\lambda^2\,\frac{1}{2}\,\partial_\theta
\Bigl(
\EE(p_\sigma^2)\,\partial_\theta\rho_\lambda
\,+\,
\EE( p_\sigma\partial_\theta p_\sigma)\,
\rho_\lambda
\,-\,
2\,\EE(q_\sigma)\,\rho_\lambda
\Bigr)
\,+\,
\Oo(\lambda^3)
\,=\,
\rho_\lambda
\,.
$$

We first consider an anomaly of first degree. As $\EE(P_\sigma)\neq 0$, it
follows that $\EE(p_\sigma)$ is not vanishing identically. Therefore the
above perturbative equation is non-trivial to first order in $\lambda$,
hence $\EE(p_\sigma)\rho_0$ should be constant. If now the first
degree anomaly is elliptic, then $\det \EE(P_\sigma)>0$ which can easily be
seen to be equivalent to $|\EE(\alpha_\sigma)|>|\EE(\beta_\sigma)|$, which
in turn is equivalent to the fact that $\EE(p_\sigma(\theta))$ does not
vanish for any $\theta\in S^1$. For an elliptic anomaly of first degree,
the lowest order of the invariant measure is therefore

$$
\rho_0
\;=\;
\frac{c}{\EE(p_\sigma)}
\;,
$$

\noindent with an adequate normalization constant $c\in\RR$. If on the
other hand, the anomaly is hyperbolic (resp. parabolic), then
$\EE(p_\sigma)$ has four (resp. two) zeros on $S^1$. In this situation, the
only possible (formal) solution is that $\rho_0$ is given by Dirac peaks
on these zeros (which is, of course, only formal because the invariant
measure is known to be H\"older continuous). In Section \ref{sec-hyp}, we
shall see that for the calculation of certain expectation values w.r.t. the
invariant measure, it looks as if it were given by a sum of two Dirac
peaks,
concentrated on the stable fixed points of the averaged phase shift
dynamics. These fixed points are two of the zeros of $\EE(p_\sigma)$.

\vspace{.2cm}

Next we consider an anomaly of second degree.
As then $\EE(p_\sigma)=0$, it follows that the equation for the lowest
order of the invariant measure is

$$
\frac{1}{2}\,
\partial_\theta\,
\Bigl(\EE(p_\sigma^2)\,\partial_\theta\rho_0
\,+\,
\EE( p_\sigma\partial_\theta p_\sigma)\,\rho_0
\,-\,
2\,\EE(q_\sigma)\,\rho_0
\Bigr)
\;=\;
0\;.
$$

\noindent Now
$\EE(p_\sigma^2)>0$ unless $\pp$-almost all $p_\sigma$ vanish
simultaneously for some $\theta$, a (rare) situation which is excluded
throughout the present work. Then this is an analytic 
Fokker-Planck equation on the unit circle and it can be written as
 $\Ll\,\rho_0=0$ where $\Ll$ is by
definition the Fokker-Planck operator. Its
spectrum contains the simple eigenvalue $0$ with eigenvector
given by the (lowest order of the) 
invariant measure $\rho_0$ calculated next. Indeed,

$$
\frac{1}{2}\,\EE(p_\sigma^2)\,\partial_\theta\rho_0
\,+\,
\frac{1}{2}\,\EE( p_\sigma\partial_\theta p_\sigma)\,\rho_0
\,-\,
\EE(q_\sigma)\,\rho_0
\;=\;
C\;,
$$

\noindent where the real constant $C$ has to be chosen such that the equation
admits a positive, $2\pi$-periodic and normalized
solution $\rho_0$. It is a routine
calculation to determine the solution using the method of variation of the
constants. Setting

$$
\kappa(\theta)
\;=\;
\int^\theta_0 d\theta'\,\frac{2\,\EE(q_\sigma(\theta'))
}{\EE(p_\sigma^2(\theta'))}
\;,
\qquad
K(\theta)
\;=\;
\int^\theta_0
d\theta'\,2\,\EE(p_\sigma^2(\theta'))^{-\frac{1}{2}}\,e^{-\kappa(\theta')}
\;,
\qquad
C
\;=\;
\frac{e^{-\kappa(2\pi)}-1}{K(2\pi)}
\;,
$$

\noindent it is given by

\begin{equation}
\label{eq-invariantdens}
\rho_0(\theta)
\;=\;
\frac{c\,e^{\kappa(\theta)}}{\EE(p_\sigma^2(\theta))^{\frac{1}{2}}}
\left(C\,K(\theta)\,+\,1
\right)
\;,
\end{equation}

\noindent where $c$ is a normalization constant. It is important to
note at this point that $\rho_0(\theta)$ 
is an analytic function of $\theta$. 

\vspace{.2cm}

The rest of the
spectrum of $\Ll$ is discrete ($\Ll$ has a compact resolvent), at most
twice
degenerate and has a strictly negative real part, all facts that can be
 proven as indicated in \cite{Ris}. As already stated in the introduction,
we do not need to use this spectral information directly.

%%%%%%%%%%%%%%%%%%%%%%%%%%%%%%%%%%%%%%%%%%
\section{The Lyapunov exponent}

The asymptotic behavior of the products of the random sequence
of matrices $(T_{\lambda,\sigma_n})_{n\geq 1}$
is characterized by the Lyapunov exponent \cite[A.III.3.4]{BL}

\begin{equation}
\label{eq-lyap}
\gamma(\lambda)
\;=\;
\lim_{N\to\infty}
\frac{1}{N}
\,\EE\,\log\left(\left\|\prod_{n=1}^N \,T_{\lambda,\sigma_n}
\;e_\theta
\right\|\right)
\mbox{ , }
\end{equation}

\noindent where $\theta$ is an arbitrary initial condition. One may
also average over $\theta$ w.r.t. an arbitrary continuous measure
before taking the limit \cite[Lemma~3]{JSS}.
A result of Furstenberg states a criterion for having a positive
Lyapunov exponent \cite{BL}. A quantitative control of the Lyapunov
exponent in the vicinity of a
critical point is given in \cite[Proposition~1]{SSS}, however, only in
the case where the critical point is not an anomaly of first or second
order. The latter two cases are dealt with in the present work.

\vspace{.2cm}

Let us first suppose that the anomaly is of first order.
Because the boundary terms vanish in the limit, it is possible to use the
matrices $MT_{\lambda,\sigma_n}M^{-1}$ instead of $T_{\lambda,\sigma_n}$
in (\ref{eq-lyap}). Furthermore, the random dynamical system
(\ref{eq-randomdyn}) allows to expand (\ref{eq-lyap}) into a telescopic
sum:

\begin{equation}
\label{eq-lyap2}
\gamma(\lambda)
\;=\;
\lim_{N\to\infty}
\frac{1}{N}
\;\sum_{n=0}^{N-1}
\,\EE\,\log\left(\left\| MT_{\lambda,\sigma_{n+1}}M^{-1}
\,e_{\Ss^{n}_{\lambda,\omega}(\theta)}
\right\|\right)
\mbox{ . }
\end{equation}

\noindent Up to terms of order $\Oo(\lambda^3)$, we can expand each
contribution of this Birkhoff sum:

\begin{eqnarray*}
\log\left(\left\| MT_{\lambda,\sigma}M^{-1}
\,e_\theta
\right\|\right)\!\!\!
& = & \!\!\!
\lambda\,
\langle e_\theta|P_\sigma|e_\theta\rangle
\;+\;
\frac{1}{2}\;\lambda^2
\Bigl(
\langle e_\theta|(|P_\sigma|^2+Q_\sigma+P_\sigma^2) |e_\theta\rangle
\;-\;
2\;\langle e_\theta|P_\sigma|e_\theta\rangle^2
\Bigr)
\\
& = & \!\!\!
\frac{1}{2}\,
\Re e
\left[
2\,\lambda\,\beta_\sigma\, e^{2\imath \theta}
\,+\,
\lambda^2
\Bigl(
|\beta_\sigma|^2
\,+\,
\langle \overline{v}|(|P_\sigma|^2+2\,Q_\sigma)|v\rangle
e^{2\imath \theta}\,-\,
\beta_\sigma^2\, e^{4\imath \theta}
\Bigr)
\right]
,
\end{eqnarray*}

\noindent where we used the identity

$$
\langle e_\theta|T|e_\theta\rangle
\;=\;
\frac{1}{2}\,\Tr(T)
\;+\;
\Re e
\left(
\langle \overline{v}|T|v\rangle
\, e^{2\imath \theta}
\right)
\;,
$$

\noindent holding for any real matrix $T$, as well as 
$\Tr(P_\sigma)=\Tr(Q_\sigma)=0$ and 
$\Tr(|P_\sigma|^2+P_\sigma^2)=4\,|\beta_\sigma|^2$.
Let us set

\begin{equation}
\label{eq-osci}
I_j(N)
\;=\;
\frac{1}{N}
\;\sum_{n=0}^{N-1}
\,\EE
\bigl(e^{2\imath j\Ss^{n}_{\lambda,\omega}(\theta)}\bigr)
\;,
\qquad
j=1,2
\;,
\end{equation}

\noindent and introduce $I_j$ by $I_j(N)=I_j+\Oo(\lambda)$ for $N$
sufficiently large. 
We therefore obtain for an anomaly of first order

\begin{equation}
\label{eq-lyapexpan}
\gamma(\lambda)
\;=\;
\frac{1}{2}\;
\EE\;\Re e
\left(
2\,\lambda\,\beta_\sigma\,I_1
\;+\;
\lambda^2\,
\Bigl(
|\beta_\sigma|^2
\;+\;
\langle \overline{v}|(|P_\sigma|^2+2\,Q_\sigma)|v\rangle
\;I_1
\;-\;
\beta_\sigma^2\, I_2
\Bigr)
\right)
\;+\;
\Oo(\lambda^3)
\;.
\end{equation}

For an anomaly of second order, one regroups the contributions pairwise as
in Definition~\ref{def-critical}, namely works with the family
$(T_{\lambda,\hat{\sigma}})_{\lambda\in\RR,\hat{\sigma}\in{\Sigma}^2}$
where
$T_{\lambda,\hat{\sigma}}=T_{\lambda,\sigma(2)}T_{\lambda,\sigma(1)}$
for $\hat{\sigma}=(\sigma(2),\sigma(1))$, furnished with the probability
measure $\hat{\pp}=\pp^{\times 2}$. This family has an anomaly of first
order, and its Lyapunov exponent is exactly twice that of the initial
family $(T_{\lambda,\sigma})_
{\lambda\in\RR,\sigma\in\Sigma }$. It is hence sufficient to study 
anomalies of first order. 

%%%%%%%%%%%%%%%%%%%%%%%%%%%%%%%%%%%%%%%%%%
\subsection{Elliptic first degree anomaly}
\label{sec-elliptic}

Let us consider the matrix $\EE(\partial_\lambda
T_{\lambda,\sigma}|_{\lambda=0})\in\,$sl$(2,\RR)$. For an elliptic anomaly,
the determinant of this matrix is positive. Its eigenvalues are therefore a
complex conjugate pair $\pm\,\imath \,\frac{\eta}{2}$, so that there exists
a basis change $M\in\,$SL$(2,\RR)$ such that

\begin{equation}
\label{eq-elliptic}
\EE\bigl(
M\partial_\lambda
T_{\lambda,\sigma}|_{\lambda=0}M^{-1}
\bigr)
\;=\;
\EE(P_\sigma)
\;=\;
\frac{1}{2}\;
\left(
\begin{array}{cc}
0 & -\eta \\
\eta & 0
\end{array}
\right)
\;.
\end{equation}

\noindent It follows that $\EE(\alpha_\sigma)=\imath\,\frac{\eta}{2}$ and
$\EE(\beta_\sigma)=0$, implying $\gamma(\lambda)=
\Oo(\lambda^2)$. Furthermore, from (\ref{eq-polydef}),

$$
e^{2\imath j \Ss_{\lambda,\sigma}(\theta)}
\;=\;
\bigl(1\,+\,2\imath j\,\lambda\;\Im m(\alpha_\sigma
\,-\,\beta_\sigma\,e^{2\imath\theta})\bigr)
\;
e^{2\imath j \theta}
\;+\;
\Oo(\lambda^2)
\;.
$$

\noindent Hence

$$
I_j(N)
\;=\;
\frac{1}{N}
\;\sum_{n=0}^{N-1}
\,(1+\imath j \lambda\eta)\,\EE
\bigl(
e^{2\imath j\Ss^{n-1}_{\lambda,\omega}(\theta)}
\bigr)
\;+\;
\Oo(\lambda^2)
\;=\;
(1+\imath j \lambda\eta)\;I_j(N)\;+\;
\Oo\bigl(N^{-1},\lambda^2\bigr)
\,,
$$

\noindent so that
$I_{j}(N)=\Oo(\lambda,(N\lambda)^{-1})$ and
$I_j=0$. Replacing into (\ref{eq-lyapexpan}),
this leads to:

%%%%%%%%%%%%%%%%%%%%%%%%%%%
\begin{proposi}
\label{prop-firstell}
If $\lambda=0$ is an elliptic anomaly of first order and first degree,
then
$$
\gamma(\lambda)
\;=\;
\frac{1}{2}\,
\lambda^2\;
\EE(|\beta_\sigma|^2)
\;+\;
\Oo(\lambda^3)
\;.
$$

\end{proposi}
%%%%%%%%%%%%%%%%%%%%%%%%%%%

In order to calculate $\beta_\sigma$ in an application, one
first has to determine the basis change (\ref{eq-elliptic}), then
$P_\sigma$ before deducing $\beta_\sigma$ and $\EE(|\beta_\sigma|^2)$, see
Section~\ref{sec-examples} for two examples.
Let us note that after the basis change, (\ref{eq-elliptic}) implies
$\EE(p_\sigma(\theta))=\eta$. Hence the invariant measure is to lowest
order given by the Lebesgue measure after the basis change.

\vspace{.2cm}

The term 'elliptic' indicates that the mean dynamics at the anomaly is to
lowest order a rotation. For a hyperbolic anomaly it is an expansion in a
given direction and a contraction into another one. These directions ({\it
i.e.} angles) can be chosen to our convenience through the basis
change $M$, as will be done next.

%%%%%%%%%%%%%%%%%%%%%%%%%%%%%%%%%%%%%%%%%%
\subsection{Hyperbolic first degree anomaly}
\label{sec-hyp}

For a hyperbolic anomaly of first degree, the eigenvalues of 
$\EE(\partial_\lambda
T_{\lambda,\sigma}|_{\lambda=0})$ are 
$\pm\frac{\mu}{2}$ and there exists $M\in\,$SL$(2,\RR)$ such that

\begin{equation}
\label{eq-hyperbolic}
\EE\bigl(
M\partial_\lambda
T_{\lambda,\sigma}|_{\lambda=0}M^{-1}
\bigr)
\;=\;
\EE(P_\sigma)
\;=\;
\frac{1}{2}\;
\left(
\begin{array}{cc}
\mu & 0 \\
0 & -\mu
\end{array}
\right)
\;.
\end{equation}

\noindent It follows that $\EE(\alpha_\sigma)=0$ and
$\EE(\beta_\sigma)=\frac{\mu}{2}$, so that
(\ref{eq-lyapexpan}) leads to

$$
\gamma(\lambda)
\;=\;
\frac{1}{2}\;
\lambda\,\mu\;\Re e(I_1)
\;+\;
\Oo(\lambda^2)
\;.
$$ 

\noindent We now need to evaluate $I_1$. Introducing the reference
dynamics 
$\tilde{\Ss}_{\lambda}(\theta)=\theta-\frac{1}{2}
\lambda\mu\sin(2\theta)$ as well as the
centered perturbation $r_\sigma(\theta)=\Im m(\alpha_\sigma
-(\beta_\sigma-\frac{\mu}{2})\,e^{2\imath\theta})$, it
follows from (\ref{eq-polydef}) that the phase shift dynamics 
is 

\begin{equation}
\label{eq-centerdyn}
\Ss_{\lambda,\sigma}(\theta)
\;=\;
\tilde{\Ss}_{\lambda}(\theta)
\;+\;
\lambda\,
r_\sigma(\theta)
\;+\;
\Oo(\lambda^2)
\;.
\end{equation}

\noindent The non-random dynamics $\tilde{\Ss}_\lambda$ has four fixed
points, $\theta=0,\frac{\pi}{2},\pi,\frac{3\pi}{2}$. 
If $\lambda\mu>0$, $\theta=0$ and $\pi$ are
stable, while $\theta=\frac{\pi}{2}$ and $\frac{3\pi}{2}$ are 
unstable. For $\lambda\mu<0$, the roles are
exchanged, and this case will not be considered here. Unless the initial
condition is the unstable fixed point, one has
$\tilde{\Ss}^n_\lambda(\theta)\to 0,\pi$ as $n\to\infty$. Furthermore, if
$\theta$ is not within a $\Oo({\lambda}^{\frac{1}{2}})$-neighborhood
of an unstable fixed point, it takes $n=\Oo(\lambda^{-\frac{3}{2}})$
iterations of
$\tilde{\Ss}_\lambda$ in order to attain a
$\Oo({\lambda}^{\frac{1}{2}})$-neighborhood of $0$.
We also need to expand iterations of $\hat{\Ss}_\lambda$:

$$
\tilde{\Ss}^k_\lambda\bigl(
\theta\,+\,\lambda\,r_\sigma(\theta')\,+\,
\Oo(\lambda^2)\bigr)
\;=\;
\tilde{\Ss}^k_\lambda(\theta)
\;+\;
\lambda\,\partial_\theta\tilde{\Ss}^k_\lambda(\theta)\,
r_\sigma(\theta')\;+\;
\Oo(\lambda^2)
\;,
$$

\noindent where the corrective term $\Oo(\lambda^2)$ on the r.h.s. is
bounded uniformly in $k$ as one readily realizes when thinking of the
dynamics induced by $\tilde{\Ss}_\lambda$. Furthermore, 
$\partial_\theta\tilde{\Ss}^k_\lambda(\theta)=\Oo(1)$ uniformly in $k$.
Iteration thus shows:

$$
\Ss^n_{\lambda,\omega}(\theta)
\;=\;
\tilde{\Ss}^n_{\lambda}(\theta)
\;+\;\lambda\,
\sum_{k=1}^{n}
\partial_\theta\tilde{\Ss}^{n-k}_\lambda
\bigl( \tilde{\Ss}_\lambda(\Ss^{k-1}_{\lambda,\omega}(\theta))\bigr)
\;
r_{\sigma_{k}}(\Ss^{k-1}_{\lambda,\omega}(\theta))
\;+\;
\Oo(n\lambda^2)
\;.
$$

\noindent Let us denote the coefficient in the sum over $k$ by $s_k$. Then
$s_k$ is a random variable that depends only on $\sigma_l$ for $l\leq k$.
Moreover, $s_k$ is centered when a conditional expectation over $\sigma_k$
is taken. Taking successively conditional expectations thus shows

\begin{equation}
\label{eq-help0}
\EE
\bigl(
e^{2\imath
\Ss^n_{\lambda,\omega}(\theta)}\bigr)
\;=\;
e^{2\imath
\tilde{\Ss}^n_{\lambda}(\theta)}
\;\Bigl(\EE\bigl(
e^{2\imath\lambda\sum_{k=1}^{n-1}s_k
\,+\,
\Oo(n\lambda^2)}
\bigr)
\;+\;
\Oo(\lambda^2)
\Bigr)
\;=\;
e^{2\imath
\tilde{\Ss}^n_{\lambda}(\theta)}
\;+\;
\Oo(n\lambda^2)
\;.
\end{equation}

\noindent Choosing $n=\lambda^{-\frac{3}{2}}$ gives according to the
above

\begin{equation}
\label{eq-help}
\EE
\bigl(
e^{2\imath
\Ss^n_{\lambda,\omega}(\theta)}\bigr)
\;=\;
1
\;+\;
\Oo(\lambda^{\frac{1}{2}})
\;,
\end{equation}

\noindent unless $\theta$ is within a 
$\Oo({\lambda}^{\frac{1}{2}})$-neighborhood
of $\frac{\pi}{2}$ or $\frac{3\pi}{2}$. 
In the latter cases, an elementary argument based on the central
limit theorem shows that it takes of order 
$\EE(r_\sigma(\frac{\pi}{2})^2){\lambda}^{-\frac{3}{2}}$ iterations to diffuse
out of these regions left out before. Supposing that one does
not have $r_\sigma(\frac{\pi}{2})= 0$
for $\pp$-almost all $\sigma$, one can conclude that
(\ref{eq-help}) holds
for all initial conditions $\theta$. Consequently
$I_1=1+\Oo(\lambda^{\frac{1}{2}})$ so that:

%%%%%%%%%%%%%%%%%%%%%%%%%%%
\begin{proposi}
\label{prop-firsthyp}
If $\lambda=0$ is an hyperbolic anomaly of first order and first degree,
and $r_\sigma(\frac{\pi}{2})$ does not vanish
for $\pp$-almost all $\sigma$, one has

$$
\gamma(\lambda)
\;=\;
\frac{1}{2}\;
|\lambda\,\mu|
\;+\;
\Oo(\lambda^{\frac{3}{2}})
\;.
$$

\end{proposi}
%%%%%%%%%%%%%%%%%%%%%%%%%%%

The argument above shows that the random phase
dynamics is such that the angles $\Ss^n_{\lambda,\omega}(\theta)$ are for
most $n$ and $\omega$ in a neighborhood of size ${\lambda}^{\frac{1}{2}}$ 
of the stable fixed points $\theta=0,\pi$. This does not mean that for some
$n$
and $\omega$, the angles are elsewhere; in particular, the rotation number
of the dynamics does not vanish. However, this leads to corrections which
do not enter into the lowest order term for the Lyapunov exponent.

%%%%%%%%%%%%%%%%%%%%%%%%%%%%%%%%%%%%%%%%%%
\subsection{Parabolic first degree anomaly}

This may seem like a pathological and exceptional case. It turns out to be
the mathematically most interesting anomaly of first degree, though, and
its analysis is similar to that of the Lyapunov exponent near band
edges (this will be discussed elsewhere).
First of all, at a parabolic anomaly of first degree, 
there exists $M\in\,$SL$(2,\RR)$ allowing to attain the Jordan normal
form: 

\begin{equation}
\label{eq-parbolic}
\EE\bigl(
M\partial_\lambda
T_{\lambda,\sigma}|_{\lambda=0}M^{-1}
\bigr)
\;=\;
\EE(P_\sigma)
\;=\;
\left(
\begin{array}{cc}
0 & 1 \\
0 & 0
\end{array}
\right)
\;.
\end{equation}

\noindent Thus $\EE(\alpha_\sigma)=\EE(\beta_\sigma)=-\imath
\,\frac{1}{2}$ so that $\gamma(\lambda)=
\frac{1}{2}\,\lambda\,\Im m(I_1)
\,+\,\Oo(\lambda^2)$. Introducing the reference dynamics 
$\hat{\Ss}_{\lambda}(\theta)=\theta+\frac{\lambda}{2}(\cos(2\theta)-1)$
as well as the
centered perturbation $r_\sigma(\theta)=\Im
m(\alpha_\sigma+\frac{\imath}{2}
-(\beta_\sigma+\frac{\imath}{2})\,e^{2\imath\theta})$, the dynamics can
then be decomposed as in (\ref{eq-centerdyn}). Moreover, the
argument leading to (\ref{eq-help0}) directly transposes to the present
case. 
However, this does not allow to calculate the leading order
contribution, but shows that $\gamma(\lambda)=\Oo(\lambda^{\frac{3}{2}})$.

%%%%%%%%%%%%%%%%%%%%%%%%%%%%%%%%%%%%%%%%%%
\subsection{Second degree anomaly}
\label{sec-second}

At a second degree anomaly one has $\EE(\beta_\sigma)=0$, so in order
to calculate the lowest order of the Lyapunov exponent  one
needs to, according to (\ref{eq-lyapexpan}), evaluate $I_1$ and $I_2$.
For this purpose let us introduce an analytic 
change of variables $Z:S^1\to S^1$
using the density
$\rho_0$ given in (\ref{eq-invariantdens}):

$$
\hth
\;=\;
Z(\theta)
\;=\;
\int^\theta_0 d\theta'\;\rho_0(\theta')
\;.
$$

\noindent According to Section~\ref{sec-invariant}, one expects that
the distribution of $\hth$ is the Lebesgue measure.
We will only need to
prove that this holds perturbatively in a weak sense when integrating 
analytic functions. 

\vspace{.2cm}

We need to study the transformed dynamics 
$\hat{\Ss}_{\lambda,\sigma}= Z\circ {\Ss}_{\lambda,\sigma} \circ
Z^{-1}$ and write it again in the form:

\begin{equation}
\label{eq-polydef2}
\hat{\Ss}_{\lambda,\sigma}(\hth)
\;=\;
\hth\,+\,\lambda\,\hat{p}_\sigma(\hth)\,+\,\lambda^2\,\hat{q}_\sigma(\hth)
\,+\,\frac{1}{2}\,\lambda^2\, \hat{p}_\sigma(\hth)\,
\partial_{\hth} \,\hat{p}_\sigma(\hth)
\,+\,\Oo(\lambda^3)
\;.
\end{equation}

\noindent As, up to order $\Oo(\lambda^3)$,

$$
Z\circ \Ss_{\lambda,\sigma}(\theta)
\, = \,
Z(\theta)
\,+\,
\Bigl(
\lambda\,p_\sigma(\theta)
+\frac{1}{2}\,\lambda^2\, p_\sigma(\theta)\,\partial_\theta p_\sigma(\theta)
\,+\,
\lambda^2\,q_\sigma(\theta)
\Bigr)
\,\partial_\theta Z(\theta)
\;+\;
\frac{1}{2}\,\lambda^2\,p_\sigma^2\,\partial^2_\theta Z(\theta)
\;,
$$

\noindent one deduces from

$$
\partial_\theta Z\;=\;\rho_0\;,
\qquad
\partial_\theta^2 Z\;=\;
\frac{2}{\EE(p^2_\sigma)}\,
\Bigl(
C\,-\,\frac{1}{2}\,\EE(p_\sigma\partial_\theta p_\sigma)\,\rho_0
\,+\,\EE(q_\sigma)\,\rho_0
\Bigr)
\;,
$$

\noindent that, with $\theta=Z^{-1}(\hth)$,

$$
\hat{p}_\sigma(\hth)
\;=\;
p_\sigma(\theta)\,\rho_0(\theta)\;,
\qquad
\hat{q}_\sigma(\hth)
\;=\;
q_\sigma(\theta)\,\rho_0(\theta)\;.
$$

\noindent A short calculation shows that the expectation values satisfy

\begin{equation}
\label{eq-hatexpect}
\EE(\hat{p}_\sigma(\hth))
\;=\;
0\;,
\qquad
\EE(\hat{q}_\sigma(\hth))
\;=\;
\frac{1}{2}\,\EE\left(
\hat{p}_\sigma(\hth)\,\partial_\hth\,\hat{p}_\sigma(\hth)\right)
-C
\;,
\end{equation}

\noindent where $C$ is as in (\ref{eq-invariantdens}). 

\vspace{.2cm}

Given any analytic function $\hat{f}$ on $S^1$, let us introduce its
Fourier coefficients

$$
\hat{f}(\hth)
\;=\;
\sum_{m\in\ZZ}
\;\hat{f}_m\;
e^{\imath m \hth}
\;,
\qquad
\hat{f}_m
\;=\;
\int^{2\pi}_0 \!\frac{d\hth}{2\pi}\;
\hat{f}(\hth)\;e^{-\imath m \hat{\theta}}
\;.
$$

\noindent There exist $a,\xi>0$ such
that $\hat{f}_m\leq a\,e^{-\xi|m|}$. We are interested in

$$
\hat{I}_{\hat{f}}(N)
\;=\;
\frac{1}{N}\,\EE
\;\sum_{n=0}^{N-1}
\,
\hat{f}(\hth_n) 
\;,
$$

\noindent where, for sake of notational simplicity, we introduced
$\hth_n=\hat{\Ss}^{n}_{\lambda,\omega}(\hth)$ for iterations defined 
just as in (\ref{eq-randomdyn}).

%%%%%%%%%%%%%%%%%%%%%%%%%%%%%%%%%%%%%%%%%%%%%%%%%%%%
\begin{lemma}
Suppose $\EE(p_\sigma^2)>0$ and that $\hat{f}$ is analytic. Then

$$
\hat{I}_{\hat{f}}(N)
\;=\;
\hat{f}_0\,+\,\Oo(\lambda,(\lambda^2N)^{-1})
\;,
$$

\noindent with an error that depends on $\hat{f}$.
\end{lemma}
%%%%%%%%%%%%%%%%%%%%%%%%%%%%%%%%%%%%%%%%%%%%%%%%%%%%

\noindent {\bf Proof.} Set
$\hat{r}=\frac{1}{2}\EE(\hat{p}_\sigma^2)$. This is an analytic
function which is strictly positive on $S^1$. Furthermore let $\hat{F}$
be an auxilliary analytic function with Fourier coefficients
$\hat{F}_m$. Then, using (\ref{eq-polydef2}) and the 
identities (\ref{eq-hatexpect}), 

\begin{eqnarray*}
\hat{I}_{\hat{F}}(N)
\!\! & = &\!\!
\frac{1}{N}\,\EE
\;\sum_{n=0}^{N-1}
\sum_{m\in\ZZ}\hat{F}_m\,
e^{\imath m\hth_n}
\left(
1-\imath mC\lambda^2
+m\lambda^2(\imath\partial_\hth-m)\hat{r}(\hth_n)
+\Oo(m^3\lambda^3,N^{-1})
\right)
\\
& = &\!\!
\hat{I}_{\hat{F}}(N)
\;+\;
\lambda^2\,
\hat{I}_{-C\hat{F}'+\hat{r}'\hat{F}'+\hat{r}\hat{F}''}(N)
\,+\,\Oo(\lambda^3,N^{-1})
\;,
\end{eqnarray*}

\noindent where the prime denotes the derivative. Hence we deduce that for
any analytic function $\hat{F}$

\begin{equation}
\label{eq-bound}
\hat{I}_{-C\hat{F}'+(\hat{r}\hat{F}')'}(N)
\;=\;
\Oo(\lambda,(\lambda^2N)^{-1})
\;.
\end{equation}

Now,  extracting the constant term, it is clearly sufficient to show 
$\hat{I}_{\hat{f}}(N)=\Oo(\lambda,(\lambda^2N)^{-1})$ for an analytic
function
$\hat{f}$ with $\hat{f}_0=0$. But for such an $\hat{f}$ one can solve
the equation

\begin{equation}
\label{eq-ODE}
\hat{f}
\;=\;
-\,C\hat{F}'\,+\,(\hat{r}\hat{F}')'
\end{equation}

\noindent for an analytic and periodic function $\hat{F}$ and then conclude
due to (\ref{eq-bound}). Indeed, by the
method of variation of constants one can always solve (\ref{eq-ODE})
for an analytic $\hat{F}'$. This then has an antiderivative $\hat{F}$ as long
as $\hat{F}'$ does not have a constant term, {\it i.e.} the
zeroth order Fourier coefficient of the solution of (\ref{eq-ODE})
vanishes. Integrating (\ref{eq-ODE})
w.r.t. $\hth$, one sees that
this is precisely the case when $\hat{f}_0=0$ as long as
$C\neq 0$. If on the other hand $C=0$, then (\ref{eq-ODE}) can be
integrated once, and the antiderivative $\int \hat{f}$ of $\hat{f}$ chosen
such that $\hat{r}^{-1}\int \hat{f}$ does not have a constant term. Then a
second antiderivative can be taken, giving the desired function $\hat{F}$ in
this case.
\hfill $\Box$

\vspace{.2cm}

In order to use this result for the evaluation of $I_j(N)$ defined in
(\ref{eq-osci}), let us note
that 

$$
I_j(N)
\;=\;
\frac{1}{N}\,\EE
\;\sum_{n=0}^{N-1}
\;e^{2\imath j Z^{-1}(\hth_n)}
\;.
$$

\noindent Hence up to corrections the result is given by the zeroth order
Fourier coefficient of the analytic function $\hat{f}(\hth)=e^{2\imath j
  Z^{-1}(\hth)}$, so that after a change of variables one gets:

$$
I_j(N)
\;=\;
\int^{2\pi}_0 \!\frac{d\theta}{2\pi}\;
\rho_0(\theta)\;e^{2\imath j \theta}
\;+\;
\Oo(\lambda,(\lambda^2N)^{-1})
\;.
$$

%%%%%%%%%%%%%%%%%%%%%%%%%%%
\begin{proposi}
\label{prop-second}
If $\lambda=0$ is an anomaly of first order and second degree and
$\EE(p_\sigma^2)>0$, then one has, with $\rho_0$ given by
{\rm (\ref{eq-invariantdens})},
$$
\gamma(\lambda)
=
\frac{\lambda^2}{2}\,\Re e
\int^{2\pi}_0 \!\frac{d\theta}{2\pi}\;
\rho_0(\theta)
\,\left[
\EE(|\beta_\sigma|^2)
\,+\,
\EE (\langle
\overline{v}|(|P_\sigma|^2+2\,Q_\sigma)|v\rangle)\;e^{2\imath \theta}
-\EE (\beta_\sigma^2)\,e^{4\imath \theta}
\right]\,+\,
\Oo(\lambda^3).
$$

\end{proposi}
%%%%%%%%%%%%%%%%%%%%%%%%%%%

In the above, 
anomalies of first degree were classified into elliptic, hyperbolic
and parabolic. Second degree anomalies should be called 'diffusive'
(strictly if $\EE(p_\sigma^2)>0$). The random dynamics of the phases is
diffusive on $S^1$, with a varying diffusion coefficient and furthermore
submitted to a mean drift, also varying with the position. It does not
seem possible to transform this complex situation into a simple normal
form by an adequate basis change $M$.

\vspace{.2cm}

Let us note that
the Lyapunov exponent at an anomaly does depend on the higher
order term $Q_\sigma$ in the expansion (\ref{eq-expan}), while away
from an anomaly it does not depend on $Q_\sigma$ \cite{SSS}. Of course,
the coefficient of $\lambda^2$ in Proposition~\ref{prop-second} cannot be 
negative. Up to
now, no general argument could be found showing this directly  (a problem
that was solved in \cite[Proposition~1]{SSS} away from anomalies).
For this purpose, it might be of help to choose an adequate basis change
$M$.

%%%%%%%%%%%%%%%%%%%%%%%%%%%%%%%%%%%%%%%%%%
\section{Examples}
\label{sec-examples}

%%%%%%%%%%%%%%%%%%%%%%%%%%%%%%%%%%%%%%%%%%
\subsection{Center of band of the Anderson model}

The transfer matrices of the Anderson model are given by 

\begin{equation}
\label{eq-transfer}
T_{\lambda,\sigma}
\;=\;
\left(
\begin{array}{cc}
\lambda\, v_\sigma-E & -1 \\
1 & 0
\end{array}
\right)
\;,
\end{equation}

\noindent where $v_\sigma$ is a
real random variable and $E\in\RR$ is the
energy. The band center is given at $E=0$. In order to study the behavior
of the Lyapunov exponent at its vicinity, we set $E=\epsilon\lambda^2$ for
some fixed $\epsilon\in\RR$. Then the associated family of i.i.d. random
matrices has an anomaly of second order because

$$
T_{\lambda,\hat{\sigma}}
\;=\;
T_{\lambda,{\sigma}_2}T_{\lambda,{\sigma}_1}
\;=\;
-\;\exp
\left(
\;
\lambda
\left(
\begin{array}{cc}
0 & v_{\sigma_2} \\
-\,v_{\sigma_1} & 0
\end{array}
\right)
\;+\;
\lambda^2
\left(
\begin{array}{cc}
-\,\frac{1}{2}\,v_{\sigma_1} v_{\sigma_2}  & -\,\epsilon \\
\epsilon & \frac{1}{2}\,v_{\sigma_1} v_{\sigma_2} 
\end{array}
\right)
\;+\;
\Oo(\lambda^3)
\;
\right)
\;,
$$

\noindent where $\hat{\sigma}=(\sigma_2,\sigma_1)$. It follows that 
$\alpha_{\hat{\sigma}}=\frac{1}{2\imath}(v_{\sigma_2}+v_{\sigma_1})$ and
$\beta_{\hat{\sigma}}=\frac{1}{2\imath}(v_{\sigma_2}- v_{\sigma_1} )$.
If now $v_\sigma$ 
is not centered, then one has an elliptic anomaly of first degree and 
Proposition~\ref{prop-firstell}, combined with the factor $1/2$ due 
to the order of the anomaly, implies directly (no basis change needed here) that

$$
\gamma(\lambda)
\;=\;
\frac{1}{8}\,
\lambda^2\;
(\EE(v_\sigma^2)-\EE(v_\sigma)^2)
\;+\;
\Oo(\lambda^3)
\;.
$$

If, on the other hand, $v_\sigma$ is centered, one has a second degree
anomaly and can apply Proposition~\ref{prop-second}. One readily verifies
that $\EE(|\beta_{\hat{\sigma}}|^2)=\frac{1}{2} \EE(v_\sigma^2)$ and 
$\EE(\beta_{\hat{\sigma}}^2)=-\frac{1}{2} \EE(v_\sigma^2)$, and furthermore
that the second term in Proposition~\ref{prop-second} always vanishes, so
that

$$
\gamma(\lambda)
\;=\;
\frac{1}{8}\,
\lambda^2\;
\EE(v_\sigma^2)
\int^{2\pi}_0 \!\frac{d\theta}{2\pi}\;
\rho_0(\theta)\;(1\,+\,\cos(4\theta))
\;+\;
\Oo(\lambda^3)
\;,
$$

\noindent which is strictly positive unless $v_\sigma$ vanishes
identically (it was already supposed to be centered). If one wants to go
further, $\EE(|p_{\hat{\sigma}}(\theta)|^2)=\frac{1}{2} 
\EE(v_\sigma^2)(1+\cos^2(2\theta))$. In the case
$\epsilon=0$, one then has $\EE(Q_{\hat{\sigma}})=0$ so that
$\rho_0(\theta) = c(1+\cos^2(2\theta))^{-\frac{1}{2}}$
with a normalization constant $c$ that can be calculated by
a contour integration. This proves the formula given in \cite{DG}.

%%%%%%%%%%%%%%%%%%%%%%%%%%%%%%%%%%%%%%%%%%
\subsection{A particular random dimer model}
\label{sec-dimer}

In the random dimer model, the transfer matrix is given by the square of
(\ref{eq-transfer}), with a potential that can only take two values
$\lambda v_\sigma =\sigma v$ where $v\in\RR$ and $\sigma\in\{-1,1\}$ ({\it
e.g.} \cite{JSS}). Hence for a given energy $E\in\RR$ it is

$$
T_{\sigma}^E
\;=\;
\left(
\begin{array}{cc}
\sigma\, v-E & -1 \\
1 & 0
\end{array}
\right)^2
\;.
$$

\noindent Now the energy is chosen to be $E=v+\lambda$. Then 
$T_{\lambda,\sigma}=T_{\sigma}^{v+\lambda}$ has a critical point if
$|v|<2$. For the particular value $v=\frac{1}{\sqrt{2}}$ fixing hence a
special type of dimer model, one has:

$$
T_{\lambda,\sigma}
\;=\;
\left(
\begin{array}{cc}
\frac{1}{\sqrt{2}}\;(\sigma-1)\,-\,\lambda\;\; & -1 \\
1 & 0
\end{array}
\right)^2
\,=\,
\left(
\begin{array}{cc}
-\,\sigma\, -\,\sqrt{2}\;(\sigma-1)\,\lambda\;\;
&
\frac{1}{\sqrt{2}}\;(1-\sigma)\,+\,\lambda \\
\frac{1}{\sqrt{2}}\;(\sigma-1)\,-\,\lambda & -\,1 
\end{array}
\right)
\,+\,
\Oo(\lambda^2)
\,.
$$

\noindent This family has now an anomaly of second order and first degree
because

$$
(T_{\lambda,\sigma})^2
\;=\;
\sigma\,\exp
\left(
\lambda\,
\left(
\begin{array}{cc}
\sqrt{2}\;(\sigma-1)\;\;
& -\,3\,+\,\sigma \\
3\,-\,\sigma & \,-\,\sqrt{2}\;(\sigma-1)
\end{array}
\right)
\;+\;
\Oo(\lambda^2)
\right)
\;.
$$

\noindent One readily verifies that the determinant of 
$\EE(M^{-1}P_\sigma M)$ is equal to $7-2\EE(\sigma)-\EE(\sigma)^2$ and
hence
positive so that the anomaly is elliptic. Therefore the general result
of Section~\ref{sec-elliptic} can be applied. Let us set
$e=\EE(\sigma)$. The adequate basis change (without normalization of the 
determinant) is

$$
M
\;=\;
\left(
\begin{array}{cc}
\sqrt{7-2e-e^2} & 0 \\
\sqrt{2}(1-e) & 3-e
\end{array}
\right)
\;.
$$

\noindent A calculation then gives

$$
P_\sigma
\;=\;
\frac{1}{3-e}
\left(
\begin{array}{cc}
2\sqrt{2}(\sigma-e) &
(\sigma-3) \sqrt{7-2e-e^2}
\\
\frac{4(1-e)(\sigma-e)-(e-3) (7-\sigma-e-e^2)}{\sqrt{7-2e-e^2}}
& 
-2\sqrt{2}
(\sigma-e)
\end{array}
\right)
\;,
$$

\noindent allowing to extract $\beta_\sigma$ and then
$\EE(|\beta_\sigma|^2)$, leading to (this contains a factor $1/2$ because
the anomaly is of second order)

$$
\gamma(\lambda)
\;=\;
\lambda^2\;
\frac{2(1-e^2)}{(3-e)^2}\;
\left(
1+\frac{2(e-1)^2}{7-2e-e^2}
\right)
\;+\;
\Oo(\lambda^3)
\;.
$$

\noindent Note that if $e=1$ or $e=-1$ so that there is no randomness, the
coefficient vanishes. This special case was left out in
\cite{JSS} (the
condition $\EE(e^{4\imath\eta})\neq 1$ in the theorems of \cite{JSS} is
violated). Within the wide class of polymer models discussed in \cite{JSS},
models with all types of anomalies can be constructed, and
then be analyzed by the techniques of the present work.

%%%%%%%%%%%%%%%%%%%%%%%%%%

\end{document}